\documentclass{PoS}
\usepackage{epsfig}
\usepackage{graphics}
\usepackage{axodraw2}
\usepackage{amsmath,amssymb,lmodern}
\usepackage{xspace}

\def\FORM{{\sc Form}\xspace}

\newcolor{DarkYellow}{0 0 1 0.3}
\newcolor{MedGray}{0 0 0 0.25}

\def\VertexA(#1,#2){\BCirc(#1,#2){1.5}}
\def\VertexB(#1,#2){\CBoxc(#1,#2)(3,3){Orange}{Orange}}
\def\VertexC(#1,#2){\RotatedBox(#1,#2)(3,3){45}{DarkYellow}}
\def\VertexD(#1,#2){\Vertex(#1,#2){1.5}}
\def\VertexE(#1,#2){\BBoxc(#1,#2)(3,3)}
\def\VertexF(#1,#2){\FilledRotatedBox(#1,#2)(4,4){45}{Blue}}
\def\LineA(#1,#2)(#3,#4){\Line(#1,#2)(#3,#4)}
\def\LineB(#1,#2)(#3,#4){\Line[dash,dsize=1](#1,#2)(#3,#4)}
\def\LineC(#1,#2)(#3,#4){\Line[dash,dsize=2](#1,#2)(#3,#4)}
\def\LineD(#1,#2)(#3,#4){\Line[dash,dsize=3](#1,#2)(#3,#4)}
\def\LineE(#1,#2)(#3,#4){\Line[dash,dsize=4](#1,#2)(#3,#4)}
\def\LineF(#1,#2)(#3,#4){\Line(#1,#2)(#3,#4)}

\title{%%% for preprint version only:
\vspace*{-1.9cm}
\begin{minipage}{\textwidth}
{\normalfont\small Nikhef 2016-040, LTH 1096
\hspace{\fill} July 2016}\\
\end{minipage}\\[80pt]
 FORM, Diagrams and Topologies} 
\ShortTitle{FORM, Diagrams and Topologies}

\author{Franz Herzog$^{\,a}$, Ben Ruijl$^{\,ab}$, Takahiro Ueda$^{\,a}$, 
\speaker{J.A.M.  Vermaseren}$^{\,a}$, Andreas Vogt$^{\,c}$\\ \\
 \llap{$^a$} Nikhef Theory Group, Science Park 105, 1098 XG Amsterdam,
             NL\\
 \llap{$^b$} Leiden Centre of Data Science, Leiden University,
             Niels Bohrweg 1, 2333 CA Leiden, NL$\!\!\!\!\!$\\
 \llap{$^c$} Department of Mathematical Sciences, University of Liverpool,
            Liverpool L69 3BX, UK \\ \\
 E-mails: \email{fherzog@nikhef.nl},
                \email{benrl@nikhef.nl},
                \email{tueda@nikhef.nl},
                \email{t68@nikhef.nl},
                \email{Andreas.Vogt@liv.ac.uk}  \\ \\ \\
		}

\abstract{We discuss a number of FORM features that are essential in the 
automatic processing of very large numbers of diagrams as used in the 
Forcer program for 4-loop massless propagator diagrams. Most of these 
features are new.  \\  \vspace{3.5cm}}

\FullConference{ Loops and Legs in Quantum Field Theory - LL 2016,\\
		 25 April - 29 April 2014, Leipzig, Germany }

\begin{document}

%--#] Startup :
%--#[ Introduction :

\section{Introduction}

When we do a higher order calculation in field theory roughly 5 steps are 
involved:
\begin{description}
\item[Diagram generation]
\item[Notation] Bring diagrams to a notation that is compatible with the libraries.
\item[Physics] Projections, Feynman rules, expansions, rearrangements, etc.
\item[Reduction] Express everything in terms of master integrals.
\item[Integration] Work out the master integral.
\end{description}

Here we will concentrate on the second step and some aspects of the fourth 
step as encountered in the Forcer program~\cite{Forcer}. For the first step 
we use QGRAF~\cite{Qgraf}, we use Forcer in the fourth step and we take the 
master integrals from the papers by Baikov and Chetyrkin~\cite{master1}, 
and Lee, Smirnov and Smirnov~\cite{master2}. Of course we use 
\FORM~\cite{Form, Form4} throughout.

The third step is in principle trivial (unless you are into IR 
rearrangements and $R^\star$ operations), but if one likes to use gauge 
parameters and lots of gluons, it is possible to make some essential 
optimizations.

With quarks Ward identities can simplify things. For the gluons the key 
point is to notice that the canonical representation of the vertex
\begin{eqnarray}
    V^{\mu\nu\rho}(p_1,p_2,p_3) & = &
     (p_1^\rho-p_2^\rho)\ \delta^{\mu\nu}
    +(p_2^\mu-p_3^\mu)\ \delta^{\nu\rho}
    +(p_3^\nu-p_1^\nu)\ \delta^{\rho\mu}
\end{eqnarray}
is rarely the most efficient. Consider what happens when one or more 
indices are contracted with 'its own vector':
\begin{eqnarray}
    V^{\mu\nu\rho}(p_1,p_2,p_3)\ p_3^\rho & = & 
        (p_2.p_2-p_1.p_1)\ \delta^{\mu\nu}+p_1^\mu\ p_1^\nu - p_2^\mu\ p_2^\nu
          \\
    V^{\mu\nu\rho}(p_1,p_2,p_3)\ p_2^\nu p_3^\rho & = &
         p_1^\mu\ p_1.p_2 - p_2^\mu\ p_1.p_1  \\
    V^{\mu\nu\rho}(p_1,p_2,p_3)\ p_1^\mu p_2^\nu p_3^\rho & = & 0
\end{eqnarray}
You can see an obvious improvement in the number of terms and the killing 
of denominators. Additionally many terms pass vectors with 'their own 
index' on to the next vertex, leading to more simplifications. To start 
this game it is best to rewrite the triple gluon vertex as
\begin{eqnarray}
    V^{\mu\nu\rho}(p_1,p_2,p_3) & = &
     (2 p_1^\rho+ p_3^\rho)\ \delta^{\mu\nu}
    +(2 p_2^\mu + p_1^\mu)\ \delta^{\nu\rho}
    +(2 p_3^\nu + p_2^\nu)\ \delta^{\rho\mu}
\end{eqnarray}
or a different rewrite (it is not unique) in a way that three vectors go 
with their own index. Starting substitutions with one such vertex and then 
each time substituting the vertex with the largest number of 
'auto-contractions' gives a very big savings on the purely gluonic diagrams. 
When gauge parameters are involved this is even more obvious.

We will start with concentrating on step two. In past work that involved 
DIS~\cite{Paulo,DIS1,DIS3} we had a program that brought the diagrams 
to the needed notations, determined the topology, the color factors, 
symmetries, etc.  We called this program convdia.frm as opposed to 
calcdia.frm that does the steps three and four (and five). The whole was 
managed by a database program with make-like facilities called MINOS (also 
available for free from the \FORM\ website).

The first version of convdia.frm was made by Paulo Nogueira~\cite{Paulo} 
when he was staying at Nikhef for a year in the mid-1990's. After that it 
underwent many changes. Hence Paulo may not recognize any of it. The 
development inspired changes in \FORM\ as well.

Because there is no good reason to change the above setup, we kept this 
line of approach, but reprogrammed most of it. In this talk we will go over 
some of the main steps, their subtleties and some of the recent changes in 
\FORM.

Next we look at a recent function in \FORM that introduces a new concept in 
symbolic manipulation.

Finally we make some remarks about expansions in $\epsilon = 2-D/2$. 

This work is part of the HEPGAME project which is supported by the ERC 
advanced grant 320651.

%--#] Introduction :
%--#[ Insertions :

\section{Insertions}

In this step we collect the diagrams which together form sub-propagators 
as insertion graphs. This means that we will have lines in which the power 
of the denominator is not an integer. This gives great savings in the 
calculation. Take for instance the diagram

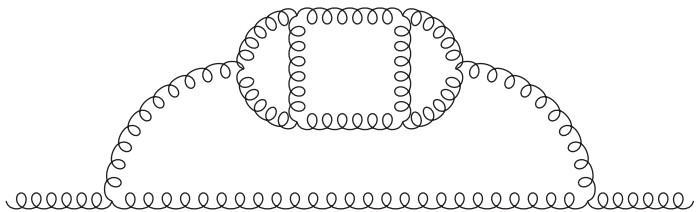
\begin{center}
\begin{axopicture}{(300,110)(0,0)}
\Gluon(130,90)(170,90){3}{6}
\GluonArc(130,70)(20,90,180){3}{5}
\GluonArc(130,70)(20,180,270){3}{5}
\GluonArc(170,70)(20,270,360){3}{5}
\GluonArc(170,70)(20,0,90){3}{5}
\Gluon(130,50)(170,50){-3}{6}
\Gluon(130,50)(130,90){3}{6}
\Gluon(170,90)(170,50){3}{6}
\GluonArc(110,20)(50,90,180){3}{10}
\GluonArc(190,20)(50,0,90){3}{10}
\Gluon(20,20)(60,20){-3}{6}
\Gluon(240,20)(280,20){-3}{6}
\Gluon(60,20)(240,20){-3}{25}
\end{axopicture}
\end{center}

This diagram is part of a one loop diagram in which one line is a three 
loop gluon propagator. Such a 3-loop propagator contains actually many 
diagrams, 254 to be precise, and is relatively expensive to compute.

In addition it comes back in other calculations as in

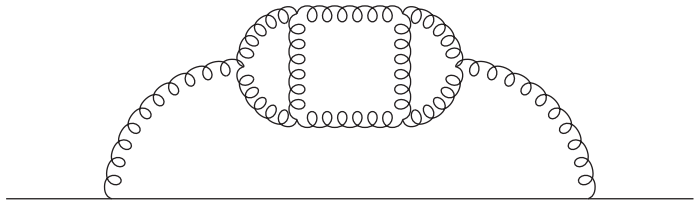
\begin{center}
\begin{axopicture}{(300,110)(0,0)}
\Gluon(130,90)(170,90){3}{6}
\GluonArc(130,70)(20,90,180){3}{5}
\GluonArc(130,70)(20,180,270){3}{5}
\GluonArc(170,70)(20,270,360){3}{5}
\GluonArc(170,70)(20,0,90){3}{5}
\Gluon(130,50)(170,50){-3}{6}
\Gluon(130,50)(130,90){3}{6}
\Gluon(170,90)(170,50){3}{6}
\GluonArc(110,20)(50,90,180){3}{10}
\GluonArc(190,20)(50,0,90){3}{10}
\Line(20,20)(60,20)
\Line(240,20)(280,20)
\Line(60,20)(240,20)
\end{axopicture}
\end{center}

\noindent
or in the computation of splitting functions and coefficient functions:

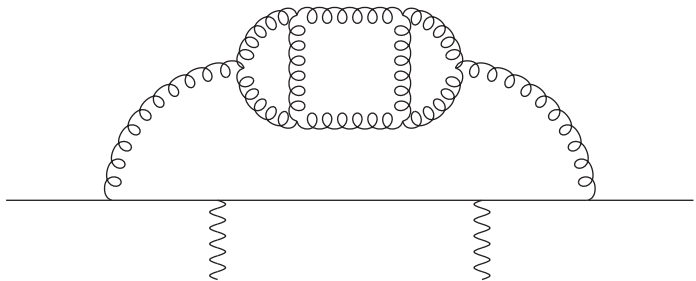
\begin{center}
\begin{axopicture}{(300,130)(0,0)}
%\AxoGrid(0,0)(10,10)(30,13){LightGray}{0.5}
\Gluon(130,110)(170,110){3}{6}
\GluonArc(130,90)(20,90,180){3}{5}
\GluonArc(130,90)(20,180,270){3}{5}
\GluonArc(170,90)(20,270,360){3}{5}
\GluonArc(170,90)(20,0,90){3}{5}
\Gluon(130,70)(170,70){-3}{6}
\Gluon(130,70)(130,110){3}{6}
\Gluon(170,110)(170,70){3}{6}
\GluonArc(110,40)(50,90,180){3}{10}
\GluonArc(190,40)(50,0,90){3}{10}
\Line(20,40)(60,40)
\Line(240,40)(280,40)
\Line(60,40)(240,40)
\Photon(100,40)(100,10){3}{5}
\Photon(200,40)(200,10){-3}{5}
\end{axopicture}
\end{center}

Hence it is a good idea to compute all possible propagators to a sufficient 
number of loops in terms of master integrals and rational coefficients. 
This way we are guaranteed to have sufficient accuracy by the time we have 
to add things. And now we have to compute only the diagrams

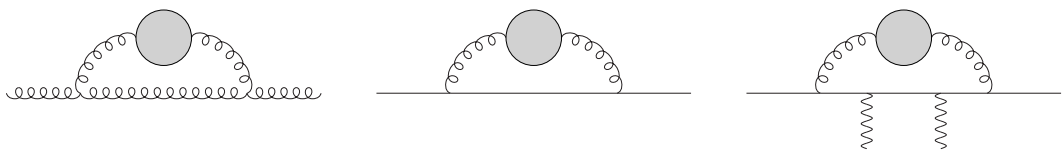
\begin{center}
\begin{axopicture}(140,84)(0,0)
\SetScale{0.7}
\GluonArc(85,40)(30,90,180){3}{5}
\GluonArc(115,40)(30,0,90){3}{5}
\GCirc(100,70){15}{0.82}
\Gluon(15,40)(55,40){-3}{5}
\Gluon(145,40)(185,40){-3}{5}
\Gluon(55,40)(145,40){-3}{12}
\end{axopicture}
\begin{axopicture}(140,84)(0,0)
\SetScale{0.7}
\GluonArc(85,40)(30,90,180){3}{5}
\GluonArc(115,40)(30,0,90){3}{5}
\GCirc(100,70){15}{0.82}
\Line(15,40)(55,40)
\Line(145,40)(185,40)
\Line(55,40)(145,40)
\end{axopicture}
\begin{axopicture}(140,84)(0,0)
\SetScale{0.7}
\GluonArc(85,40)(30,90,180){3}{5}
\GluonArc(115,40)(30,0,90){3}{5}
\GCirc(100,70){15}{0.82}
\Line(15,40)(55,40)
\Line(145,40)(185,40)
\Line(55,40)(145,40)
\Photon(80,40)(80,10){3}{5}
\Photon(120,40)(120,10){-3}{5}
\end{axopicture}
\end{center}

It does mean however that our master integrals will look a 
little bit different from the ones in the literature:

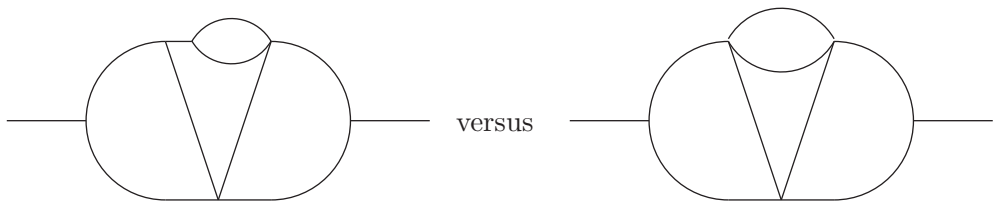
\begin{center}
\begin{axopicture}{(180,100)(0,0)}
\CArc(70,50)(30,90,270)
\CArc(110,50)(30,270,450)
\Line(10,50)(40,50)
\Line(140,50)(170,50)
\Line(90,20)(70,80)
\Line(90,20)(110,80)
\Line(70,20)(110,20)
\CArc(95,71.34)(17.32,30,150)
\CArc(95,88.86)(17.32,210,330)
\Line(70,80)(80,80)
\end{axopicture}
\raisebox{46pt}{versus}
\begin{axopicture}{(180,100)(0,0)}
\CArc(70,50)(30,90,270)
\CArc(110,50)(30,270,450)
\Line(10,50)(40,50)
\Line(140,50)(170,50)
\Line(90,20)(70,80)
\Line(90,20)(110,80)
\Line(70,20)(110,20)
\CArc(90,69.453)(23.094,30,150)
\CArc(90,91.547)(23.094,210,330)
\end{axopicture}
\end{center}
The conversion is of course simple. Put the right diagram through Forcer 
and it expresses the master from the literature in terms of our masters. Do 
this for all of them and invert the relations.

Actually one can use this technique also in the Reduze~\cite{Reduze,Reduze2} 
approach by back-substituting the complete propagator in terms of its master 
integrals into the diagram. This could save Reduze enormous amounts of 
work.

At this point we have only one problem: how to collect all diagrams that 
form a single effective propagator? It is not easy to get QGRAF to do 
this~\cite{Qgraf2}. In our case we do this in \FORM.

We first select a unique representative diagram that belongs to 
each propagator. These are for the 1-loop, 2-loop, 3-loop gluon propagators 
(don't forget the minus sign of the ghost loop):

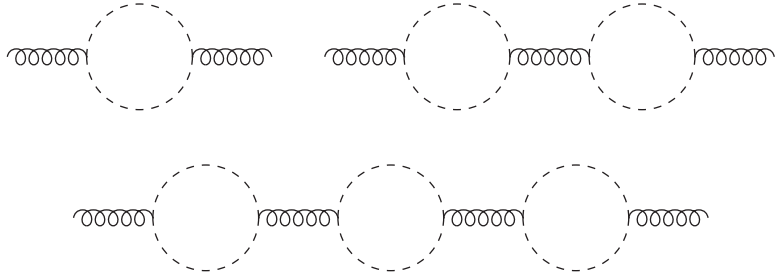
\begin{center}
\begin{axopicture}{(120,60)(0,0)}
\Gluon(10,30)(40,30){3}{5}
\DashArc(60,30)(20,180,360){3}
\DashArc(60,30)(20,0,180){3}
\Gluon(80,30)(110,30){3}{5}
\end{axopicture}
\begin{axopicture}{(190,60)(0,0)}
\Gluon(10,30)(40,30){3}{5}
\DashArc(60,30)(20,180,360){3}
\DashArc(60,30)(20,0,180){3}
\Gluon(80,30)(110,30){3}{5}
\DashArc(130,30)(20,180,360){3}
\DashArc(130,30)(20,0,180){3}
\Gluon(150,30)(180,30){3}{5}
\end{axopicture} \\
\begin{axopicture}{(260,60)(0,0)}
\Gluon(10,30)(40,30){3}{5}
\DashArc(60,30)(20,180,360){3}
\DashArc(60,30)(20,0,180){3}
\Gluon(80,30)(110,30){3}{5}
\DashArc(130,30)(20,180,360){3}
\DashArc(130,30)(20,0,180){3}
\Gluon(150,30)(180,30){3}{5}
\DashArc(200,30)(20,180,360){3}
\DashArc(200,30)(20,0,180){3}
\Gluon(220,30)(250,30){3}{5}
\end{axopicture}
\end{center}

For the quark and ghost propagators we have similar representatives.
Once we have isolated the representatives we have to throw away all other 
two-point sub-diagrams. The technical question is how to do this. In the 
intuitive way, one has to do some pretty fancy pattern matching and the 
more loops there can be, the more complicated the patterns can become and 
the larger their number. This is how it was done in the previous versions 
of convdia.

For a better solution we will make use of a new \FORM feature that was 
actually developed for something different: graph automorphisms.
We are referring to a new option for the id-statement that generates all 
different matches in the pattern matcher. This option is the "all" option.
A very simple example is:

{\footnotesize
\begin{verbatim}
    CFunction v,w;
    Vector p,p1,...,p6;
    Local F = v(p1,p2,p3)*v(p4,p5,p6);
    id,all,v(?a,p?,?b) = v(?a,p,?b)*w(p);
    Print +f +s;
    .end
   F = 
       + v(p1,p2,p3)*v(p4,p5,p6)*w(p1)
       + v(p1,p2,p3)*v(p4,p5,p6)*w(p2)
       + v(p1,p2,p3)*v(p4,p5,p6)*w(p3)
       + v(p1,p2,p3)*v(p4,p5,p6)*w(p4)
       + v(p1,p2,p3)*v(p4,p5,p6)*w(p5)
       + v(p1,p2,p3)*v(p4,p5,p6)*w(p6) ;
\end{verbatim}
}

\noindent
Now let us see how we will use that here. Assume that we have a lot of 
diagrams like
{\footnotesize
\begin{verbatim}
    +DV(glu(Q),glu(p1),glu(p2),glu(p3))
    *DV(glu(p1),GHO(p4),gho(p5))
    *DV(glu(p2),GHO(p5),gho(p6))
    *DV(glu(p7),GHO(p6),gho(p4))
    *DV(glu(p7),QUA(p8),qua(p9))
    *DV(glu(p11),QUA(p9),qua(p10))
    *DV(glu(p11),QUA(p12),qua(p8))
    *DV(glu(p13),QUA(p10),qua(p12))
    *DV(glu(Q),glu(p13),glu(p14))
    *DV(glu(p3),GHO(p15),gho(p16))*DV(glu(p17),GHO(p16),gho(p15))
    *DV(glu(p17),GHO(p18),gho(p19))*DV(glu(p14),GHO(p19),gho(p18))
\end{verbatim}
}

\noindent
which looks like
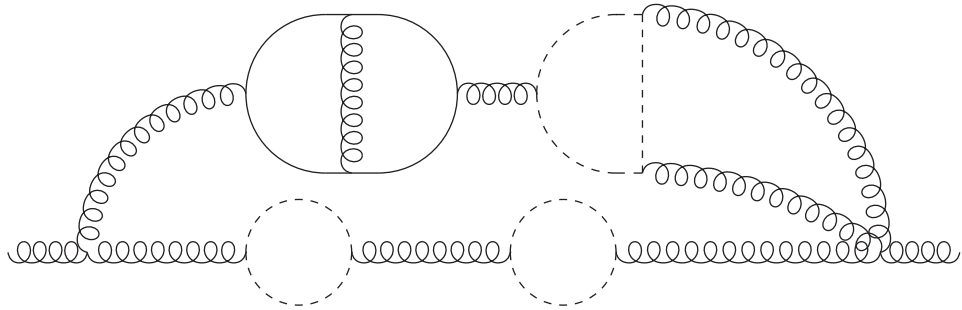
\begin{center}
\begin{axopicture}{(380,120)(0,0)}
\Gluon(10,30)(40,30){-4}{4}
\GluonArc(100,30)(60,90,180){4}{12}
\CArc(130,90)(30,90,270)
\Line(130,120)(150,120)
\Line(130,60)(150,60)
\Gluon(140,60)(140,120){4}{8}
\CArc(150,90)(30,270,450)
\Gluon(180,90)(210,90){4}{4}
\DashArc(240,90)(30,90,180){3}
\DashArc(240,90)(30,180,270){3}
\DashLine(240,120)(250,120){3}
\DashLine(240,60)(250,60){3}
\DashLine(250,60)(250,120){3}
\GluonArc(250,30)(90,0,90){4}{17}
\GluonArc(250,-90)(150,53.1301,90){4}{12}
\Gluon(340,30)(370,30){-4}{4}
\Gluon(40,30)(100,30){-4}{8}
\Gluon(140,30)(200,30){-4}{8}
\DashArc(120,30)(20,0,180){3}
\DashArc(120,30)(20,180,360){3}
\Gluon(240,30)(340,30){-4}{13}
\DashArc(220,30)(20,0,180){3}
\DashArc(220,30)(20,180,360){3}
\end{axopicture}
\end{center}
We start with taking out our representatives:
{\footnotesize
\begin{verbatim}
repeat;
*       Gluon:
id  DV(glu(p1?pp),GHO(p2?pp),gho(p3?pp))*
             DV(glu(p4?pp),GHO(p3?pp),gho(p2?pp)) =
             -dp(1,glu(p1),glu(p4))*replace_(p4,p1,p3,p0,p2,p0);
*       Ghost:
id  DV(GHO(p1?pp),gho(p2?pp),glu(p3?pp))*
             DV(gho(p4?pp),GHO(p2?pp),glu(p3?pp)) =
              dp(1,GHO(p1),gho(p4))*replace_(p4,p1,p3,p0,p2,p0);
*       Quark:
id  DV(QUA(p1?pp),qua(p2?pp),glu(p3?pp))*
             DV(qua(p4?pp),QUA(p2?pp),glu(p3?pp)) =
              dp(1,QUA(p1),qua(p4))*replace_(p4,p1,p3,p0,p2,p0);
endrepeat;
id  dp(x?,f1?(p0),f2?(p0)) = 0;
repeat id dp(x1?,?a)*dp(x2?,?a) = dp(x1+x2,?a);
\end{verbatim}
}
At this point we have a propagator function that tells how many loops are 
in it, what particles and what momentum.
Our example diagram becomes now

{\footnotesize
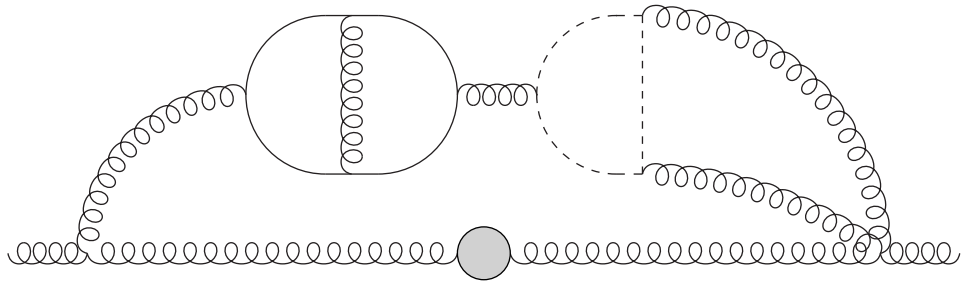
\begin{verbatim}
    +DV(glu(Q),glu(p1),glu(p2),glu(p3))
    *DV(glu(p1),GHO(p4),gho(p5))
    *DV(glu(p2),GHO(p5),gho(p6))
    *DV(glu(p7),GHO(p6),gho(p4))
    *DV(glu(p7),QUA(p8),qua(p9))
    *DV(glu(p11),QUA(p9),qua(p10))
    *DV(glu(p11),QUA(p12),qua(p8))
    *DV(glu(p13),QUA(p10),qua(p12))
    *DV(glu(Q),glu(p13),glu(p3))
    *dp(2,glu(p3),glu(p3))
\end{verbatim}
}
\begin{center}
\begin{axopicture}{(380,120)(0,0)}
\Gluon(10,30)(40,30){-4}{4}
\GluonArc(100,30)(60,90,180){4}{12}
\CArc(130,90)(30,90,270)
\Line(130,120)(150,120)
\Line(130,60)(150,60)
\Gluon(140,60)(140,120){4}{8}
\CArc(150,90)(30,270,450)
\Gluon(180,90)(210,90){4}{4}
\DashArc(240,90)(30,90,180){3}
\DashArc(240,90)(30,180,270){3}
\DashLine(240,120)(250,120){3}
\DashLine(240,60)(250,60){3}
\DashLine(250,60)(250,120){3}
\GluonArc(250,30)(90,0,90){4}{17}
\GluonArc(250,-90)(150,53.1301,90){4}{12}
\Gluon(340,30)(370,30){-4}{4}
\Gluon(40,30)(180,30){-4}{17}
\Gluon(200,30)(340,30){-4}{17}
\GCirc(190,30){10}{00.82}
\end{axopicture}
\end{center}

Now we have to eliminate all remaining diagrams that have a two-point 
function inside. For this we make a copy of our vertices in which we only 
put the momenta in the copy, and we collect these copied vertices in a 
function acc.

{\footnotesize
\begin{verbatim}
id DV(f1?(p1?),f2?(p2?),f3?(p3?)) = DV(f1(p1),f2(p2),f3(p3))*vx(p1,p2,p3);
id DV(f1?(p1?),f2?(p2?),f3?(p3?),f4?(p4?)) =
                          DV(f1(p1),f2(p2),f3(p3),f4(p4))*vx(p1,p2,p3,p4);
id  vx(?a,QQ?externals,?b) = 1;
Symmetrize vx;
id  vx(?a) = acc(vx(?a));
repeat id acc(x1?)*acc(x2?) = acc(x1*x2);
\end{verbatim}
}

Notice that we eliminate the vertices with external lines because they can 
never be part of an internal propagator. We obtain:
{\footnotesize
\begin{verbatim}
    +DV(glu(Q),glu(p1),glu(p2),glu(p3))
    *DV(glu(p1),GHO(p4),gho(p5))
    *DV(glu(p2),GHO(p5),gho(p6))
    *DV(glu(p7),GHO(p6),gho(p4))
    *DV(glu(p7),QUA(p8),qua(p9))
    *DV(glu(p11),QUA(p9),qua(p10))
    *DV(glu(p11),QUA(p12),qua(p8))
    *DV(glu(p13),QUA(p10),qua(p12))
    *DV(glu(Q),glu(p13),glu(p3))
    *dp(2,glu(p3),glu(p3))
    *acc(vx(p1,p4,p5)*vx(p2,p5,p6)*vx(p4,p6,p7)
        *vx(p7,p8,p9)*vx(p8,p11,p12)*vx(p9,p10,p11)*vx(p10,p12,p13))
\end{verbatim}
}
Now comes the algorithm: starting from each vertex, we construct the 
collection of vertices that are connected to it. We add them one by one and 
if at any time the total number of open lines of such a group is 2, this 
group is part of a propagator and we can eliminate the diagram. We have to 
try this in all possible ways.

{\footnotesize
\begin{verbatim}
$x = 0;
Argument acc;
    id,all,vx(?a) = dV(?a);
EndArgument;
Repeat;
  Argument acc;
    id,all,dV(?a,p1?,?b)*vx(?c,p1?,?d) = dV(?a,?c,?b,?d);
    Symmetrize dV;
    repeat id dV(?a,p1?,p1?,?b) = dV(?a,?b);
    if ( match(dV(p1?,p2?)) ) $x = 1;
  EndArgument;
  if ( $x == 1 ) Discard;
EndRepeat;
id  acc(x?) = 1;
\end{verbatim}
}
In the beginning we take one vx in all possible ways. This means that if 
there are 7 vertices, we generate 7 terms inside acc, each of which has one 
(different) vx replaced by dV. Then in the repeat we add, in all possible 
ways each time one vx to the dV and eliminate repeated vectors, If at any 
time any of the terms inside acc has a dV with only two arguments we set a 
mark inside \$x and the whole term can be eliminated. If this condition 
never occurs we can keep the term.

This method is amazingly versatile. It is independent of the number of 
loops inside the diagram, or inside the propagator. It is also independent 
of the number of external legs. In addition it is rather fast. We tried it 
out on the sum of all 169788 5-loop gluon propagator diagrams in the 
background gauge. With the old topology matching and some tricks by which 
we could avoid having to match 4-loop subtopologies explicitly, the 
execution time was 507 sec. With the new method it went down to 106 sec. In 
addition the code is only 38 lines instead of 127 lines. The number of 
remaining diagrams was reduced to 80064.

%--#] Insertions :
%--#[ Topologies :

\section{Topologies}

As mentioned, the id,all was actually intended for manipulating 
topologies. Here is an example of determining all automorphisms of the 
non-planar topology for 3-loop propagator graphs:
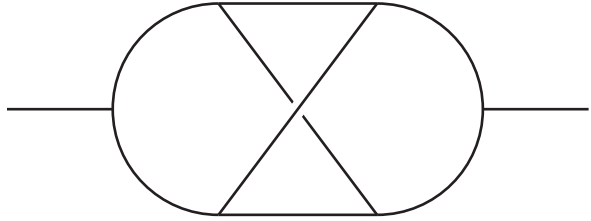
\begin{center}
\begin{axopicture}{(260,120)(0,0)}
\SetScale{2}
\SetWidth{0.5}
\Line(50,50)(80,10)
\SetColor{White}
\SetWidth{3}
\Line(50,10)(80,50)
\SetColor{Black}
\SetWidth{0.5}
\Line(50,10)(80,50)
\Line(50,50)(80,50)
\CArc(50,30)(20,90,270)
\CArc(80,30)(20,270,450)
\Line(50,10)(80,10)
\Line(10,30)(30,30)
\Line(100,30)(120,30)
\end{axopicture}
\end{center}

{\footnotesize
\begin{verbatim}
    CFunction v(s),auto;
    Vector Q,p1,...,p8,QQ,q0,...,q8;
    Set pp:<p1,-p1>,...,<p8,-p8>;
    *
    L F = v(-Q,p1,-p6)*v(-p1,p2,p7)*v(p3,-p2,p8)
         *v(-p3,p4,Q)*v(p5,-p4,-p7)*v(-p5,p6,-p8);
    $x = term_;
    id all,$x * replace_(Q,Q?{Q,-Q},<p1,p1?>,...,<p8,p8?>)
           = $x * auto(Q,QQ,<p1,q1>,...,<p8,q8>);
    *
    Bracket v;
    Print +f +s;
    .sort

   F =
    + v(-Q,-p6,p1)*v(-p1,p2,p7)*v(-p2,p3,p8)*v(-p3,Q,p4)*v(-p4,-p7,p5)*v(
   -p5,-p8,p6) * (
      + auto(Q,QQ,p1,q1,p2,q2,p3,q3,p4,q4,p5,q5,p6,q6,p7,q7,p8,q8)
      + auto(Q,QQ,p1,q1,p7,q2,-p4,q3,-p3,q4,p8,q5,p6,q6,p2,q7,p5,q8)
      + auto(Q,QQ,-p6,q1,-p5,q2,-p4,q3,-p3,q4,-p2,q5,-p1,q6,-p8,q7,-p7,q8)
      + auto(Q,QQ,-p6,q1,-p8,q2,p3,q3,p4,q4,-p7,q5,-p1,q6,-p5,q7,-p2,q8)
      + auto(-Q,QQ,p4,q1,p5,q2,p6,q3,p1,q4,p2,q5,p3,q6,-p7,q7,-p8,q8)
      + auto(-Q,QQ,p4,q1,-p7,q2,-p1,q3,-p6,q4,-p8,q5,p3,q6,p5,q7,p2,q8)
      + auto(-Q,QQ,-p3,q1,p8,q2,p6,q3,p1,q4,p7,q5,-p4,q6,-p2,q7,-p5,q8)
      + auto(-Q,QQ,-p3,q1,-p2,q2,-p1,q3,-p6,q4,-p5,q5,-p4,q6,p8,q7,p7,q8)
       );
\end{verbatim}
}
The dollar variable makes a copy of the current term and then the id 
statement matches the term onto itself, but thanks to the replace statement 
all variables have become wildcards and hence the statement generates all 
possible relabelings of the momenta. These relabelings are then stored in 
the function auto. The trick with the replace\_ in the left hand side is 
due to Takahiro Ueda. It did not work right away as intended and needed a 
considerable amount of debugging, because in the original design this was 
never seen as a possibility.
{\footnotesize
\begin{verbatim}
    id  auto(?a) = replace_(?a)/8;
    Print +f +s;
    .end

   F =
       + v(-QQ,-q6,q1)*v(-q1,q2,q7)*v(-q2,q3,q8)*v(-q3,QQ,q4)*
         v(-q4,-q7,q5)*v(-q5,-q8,q6)
      ;
\end{verbatim}
}
In the end we have the program execute those symmetry operations and we see 
that we get the input back, except for that now we have vectors q instead 
of vectors p (and QQ instead of Q).

As we saw, the id,all together with the trick to match the term with itself 
really hit the essence of graph automorphisms, and there was an immediate 
payoff in the form of the propagator example.

Now how do we determine the topology of a diagram in \FORM? And how do we 
transform to the notation that is being used in the library routines for 
those topologies? Again, in Mincer~\cite{Mince,Mince2,Mince3} this could 
still be done 'by hand' because there were basically only 18 topologies. 
Here it is different with more than 400 topologies and if one would like to 
go to 5 loops there are thousands.

The first step is that the program that (automatically) produces the 
library routines should tell us what notation it uses. In the case 
of the Forcer library this is written into a file notation.h which looks 
like (just some representatives):

{\footnotesize
\begin{verbatim}
   +vx(-Q,p4,p5)
    *vx(p3,-p4,p10)
    *vx(p2,-p3,p9)
    *vx(p1,-p2,p11)
    *vx(-p5,p6,-p11)
    *vx(-p6,p7,-p10)
    *vx(-p7,p8,-p9)
    *vx(-p1,-p8,Q)
    *TOPO(Mno1)
        ...
   +vx(-Q,p2,p3)
    *vx(p1,-p2,p5)
    *vx(-p1,p4,Q)
    *vx(-p3,-p4,-p5)
    *ex(p2,p4)
    *TOPO(Mt1star24)
\end{verbatim}
}
These are those two topologies:
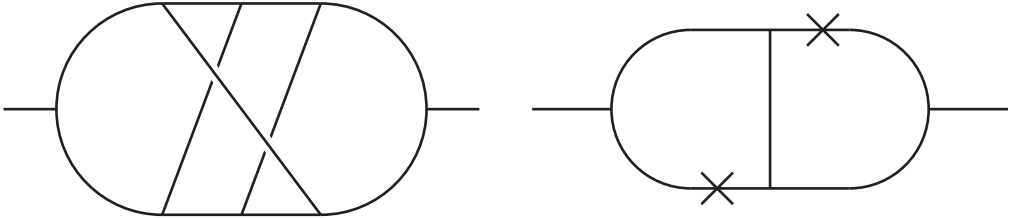
\begin{center}
\begin{axopicture}{(200,100)(0,0)}
\SetWidth{1}
\Line(70,10)(100,90)
\Line(100,10)(130,90)
\SetColor{White}
\SetWidth{5}
\Line(70,90)(130,10)
\SetColor{Black}
\SetWidth{1}
\Line(70,90)(130,10)
\CArc(70,50)(40,90,270)
\CArc(130,50)(40,270,450)
\Line(10,50)(30,50)
\Line(170,50)(190,50)
\Line(70,90)(130,90)
\Line(70,10)(130,10)
\end{axopicture}
\begin{axopicture}{(200,100)(0,0)}
\SetWidth{1}
\CArc(70,50)(30,90,270)
\CArc(130,50)(30,270,450)
\Line(10,50)(40,50)
\Line(160,50)(190,50)
\Line(70,80)(130,80)
\Line(70,20)(130,20)
\Line(100,80)(100,20)
\Line(114,74)(126,86)
\Line(114,86)(126,74)
\Line(74,14)(86,26)
\Line(74,26)(86,14)
\end{axopicture}
\end{center}
Each topology is represented by its vertices and momenta with a direction 
and a number. In addition there is a function ex of which the argument(s) 
tell us about insertions and of course the name of the topology. In the 
sequel we will ignore optimizations and just look at the main mechanism.

Step 1 is to make copies of the topologies into an array of \$-variables.
We keep there only the vertices with momenta (as in example 1) and we strip 
the signs of the momenta (for speed). We also keep the ex function but 
change it to the symmetric function EX. We 
change the function vx to the symmetric function v. We also put the names 
of the topologies (we represent them by symbols) in a set tnames.
Hence, if the first topology above becomes topology number 2 we have

{\footnotesize
\begin{verbatim}
    $topo2 = +v(Q,p4,p5)*v(p3,p4,p10)*v(p2,p3,p9)*v(p1,p2,p11)
             *v(p5,p6,p11)*v(p6,p7,p10)*v(p7,p8,p9)*v(p1,p8,Q);
    tnames[2] -> Mno1
\end{verbatim}
}

\vspace*{-1mm}
\noindent
Next we create a dictionary.
{\footnotesize
\begin{verbatim}
    #OpenDictionary wildmom
        #do i = 1,`$MAXPROPS'
            #add p`i': "p`i'?$p`i'"
        #enddo
    #CloseDictionary
\end{verbatim}
}

Originally the dictionaries were designed for creating complicated outputs. 
Their first use was for the GRACE~\cite{Grace} system. The feature we use 
here is that they print the output for a given variable as indicated in the 
text string. When this dictionary is active the system prints

{\footnotesize
\begin{verbatim}
    v(Q,p4?$p4,p5?$p5)
\end{verbatim}
}

\vspace*{-1mm}
\noindent
instead of
{\footnotesize
\begin{verbatim}
    v(Q,p4,p5)
\end{verbatim}
}
Now we are ready for the main part. We start with making a copy of all our 
vertices and propagators into the function acc. Inside acc we strip the 
particle information and bring everything to a similar notation as the 
topologies in the dollar variables. The code becomes:

{\footnotesize
\begin{verbatim}
  #do i = 1,`$MAXPROPS'
    $p`i' = p`i';
  #enddo
  $catch = 0;
  Argument acc;
    #UseDictionary wildmom($)
      #do i = 1,`$numtopo'
*      ====>
        if ( match(`$topo`i'') );
          $toponum = `i';
          $catch = 1;
          goto caught;
        endif;
      #enddo
    #CloseDictionary
    Label caught;
  EndArgument;
  if ( $catch );
    id  acc(?a) = 1;
*    ====>
    Multiply replace(Q,Q,
           <$p1,p1>,...,<$p`$MAXPROPS',p`$MAXPROPS'>)*topo($toponum);
    id  topo(n?) = topo(n,tnames[n]);
    id  replace(?a) = replace_(?a);
  elseif ( $catch == 0 );
    Print +f "The following diagram did not match a topology:\n  %t";
  endif;
\end{verbatim}
}
The key statements follow the arrows in the commentary. In the first we try 
to match a topology into the diagram. If this is successful, the values of 
the wildcards are loaded into the dollar variables \$p1,\$p2,\dots$\,$ . In the 
other statement we replace the values of the dollar variables in the 
diagram by the corresponding values in the topology.

As it is, when there are $N$ diagrams and $M$ topologies this needs on 
average $\frac{NM}{2}$ pattern matchings. This can be improved a lot by 
extra functions that tell how many 3-point and 4-point vertices there are 
and how many of those connect to an outside line. This is of course a type 
of hashing. We will skip that here.

After the topology has been determined, it is relatively easy to restore 
the signs on the momenta by comparing vertices. We skip that as well.

%--#] Topologies :
%--#[ Tables of statements :

\section{Tables of statements}

The new feature here is the id\_ function. But to explain what it does and 
why it was introduced is quite a story.

\begin{itemize}
\item Forcer uses dotproducts in the numerator, just like Mincer.
\item LiteRed~\cite{Litered,Litered2} uses a clever choice of 15 invariants.
\end{itemize}
Why do we do this differently?

\begin{itemize}
\item Reductions like the triangle rule are much more complicated in terms 
of invariants.
\item The choice of invariants or dotproducts in the numerator can 
influence the complexity of the difficult reductions very much.
\item Integrating out a one loop subdiagram is \Red{much}\ easier with 
dotproducts.
\end{itemize}
The use of the dotproducts gives much more freedom in the choice of 
variables and hence more compact reductions. (See also the talk by Takahiro 
Ueda in these proceedings~\cite{Forcer}). This goes at the cost of 
needing frequent rewritings between the various reduction steps. (You win 
some, you lose some).

Let us consider the no1 topology and what happens if the reduction removes 
a line.

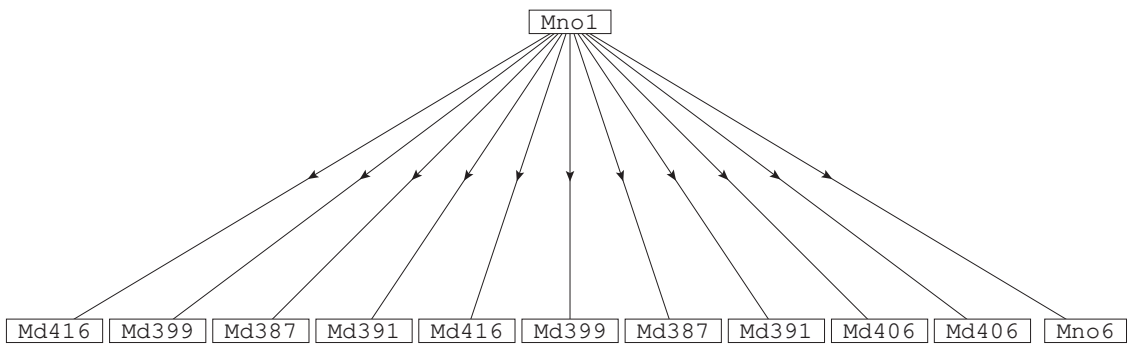
\begin{center}
\begin{axopicture}(423,130)(0,0)
%\begin{axopicture}{(650,200)(0,0)}
\SetScale{0.65}
\SetPFont{Courier}{14}
\Line[arrow,arrowpos=0.5](325,190)(25,10)
\Line[arrow,arrowpos=0.5](325,190)(85,10)
\Line[arrow,arrowpos=0.5](325,190)(145,10)
\Line[arrow,arrowpos=0.5](325,190)(205,10)
\Line[arrow,arrowpos=0.5](325,190)(265,10)
\Line[arrow,arrowpos=0.5](325,190)(325,10)
\Line[arrow,arrowpos=0.5](325,190)(385,10)
\Line[arrow,arrowpos=0.5](325,190)(445,10)
\Line[arrow,arrowpos=0.5](325,190)(505,10)
\Line[arrow,arrowpos=0.5](325,190)(565,10)
\Line[arrow,arrowpos=0.5](325,190)(625,10)
\BText(325,190){Mno1}
\BText(25,10){Md416}
\BText(85,10){Md399}
\BText(145,10){Md387}
\BText(205,10){Md391}
\BText(265,10){Md416}
\BText(325,10){Md399}
\BText(385,10){Md387}
\BText(445,10){Md391}
\BText(505,10){Md406}
\BText(565,10){Md406}
\BText(625,10){Mno6}
\end{axopicture}
\end{center}

Each of the new topologies has its own notation and its own 
'irreducible' dotproducts in the numerator. Hence there needs to be a 
relabeling. This can be done in a relatively simple way by a replace\_ 
function and in the ideal case one does this with a table. If all 
topologies have a number some table elements could look like

{\footnotesize
\begin{verbatim}
    Fill renumtable(2,0,1,1,1,1,1,1,1,1,1,1) =
           replace_(p1,Q+p7,p2,p5,p3,p3,p4,p1,p5,p2,p6,-p9,p7
           ,-p8,p8,-p7,p9,p6,p10,p4,p11,p10,Mno1,Md416);
    Fill renumtable(2,1,0,1,1,1,1,1,1,1,1,1) = ....
\end{verbatim}
}

\noindent
and after the no1 reductions have been done one can use the statement
{\footnotesize
\begin{verbatim}
    id 1/<p1.p1^n1?>/.../<p11.p11^n11?>*x?tnames[n] =
        x/<p1.p1^n1>/.../<p11.p11^n11>
        *renumtable(n,<thetap_(n1)>,...,<thetap_(n11)>);
\end{verbatim}
}
If the table is created by the (python) program that builds the whole 
system and if all table elements that are undefined are considered to be 
zero (which is the case if too many lines are missing) one needs only 
14396 table elements (and not $2^{11}\times 400+$). 

The above relabels the momenta. Now comes the tricky part. The dotproducts 
in the numerator may not be the dotproducts we need in the new topology. 
The bad part is that there can be high powers of those dotproducts when you 
want to calculate some nice large N Mellin moments in DIS~\cite{DIS1,DIS2}. 
Leaving that problem (it IS a big problem) aside we still have the case that 
we have a number of topologies that all need their numerators to be rewritten.
And we would like to do this for all of them at the same time and with as 
few pattern matchings and if statements as possible.

For the high powers with many terms on the right hand side (rhs) we want to 
do 'slow substitutions' as in

{\footnotesize
\begin{verbatim}
    id  x^n?{>=17} = x^17*(x1+...+x12)^(n-17);
    .sort
    id  x^n?{>=13} = x^13*(x1+...+x12)^(n-13);
    .sort
    id  x^n?{>=9} = x^9*(x1+...+x12)^(n-9);
    .sort
    id  x^n?{>=5} = x^5*(x1+...+x12)^(n-5);
    .sort
    id  x^n?{>=3} = x^3*(x1+...+x12)^(n-3);
    .sort
    id  x = x1+...+x12;
    .sort
\end{verbatim}
}
This way there are occasional combinations of identical terms and the whole 
process is usually much faster than a direct
\begin{verbatim}
    id  x = x1+...+x12;
    .sort
\end{verbatim}

Hence we want to set up such a system, but for each of the new topologies 
this may be different, and we will have several of such variables. One way 
would be
\begin{itemize}
\item Do this for the first topology, first variable.
\item Do this for the first topology, second variable.
\item Do this for the first topology, third variable.
\item Do this for the first topology, fourth variable.
\item Do this for the second topology, first variable.
\item etc.
\end{itemize}
This gives an enormous number of steps and in addition for each step all 
other topologies are spectators that have to be carried around. Hence we 
would like to execute this all in parallel. But the number of steps for 
each variable is not a fixed number.

To do this optimally we have a special 
function id\_. It looks a bit like replace\_ but it has extra powers:

{\footnotesize
\begin{verbatim}
    id_(p1.p2^17,p1.p2^17,p1.p2,p1.p1/2+p2.p2/2-p3.p3/2,p4.p5,
        p4.p4/2+p5.p5/2-p6.p6/2)
\end{verbatim}
}

\noindent
will be replaced during execution, after the term has been normalized, by 
the equivalent of
{\footnotesize
\begin{verbatim}
    id p1.p2^17 = p1.p2^17;
    al p1.p2 = p1.p1/2+p2.p2/2-p3.p3/2;
    al p4.p5 = p4.p4/2+p5.p5/2-p6.p6/2;
\end{verbatim}
}

\noindent
which is to say: the first 17 powers of \verb:p1.p2: are kept out from the 
further substitutions and then the other powers are replaced and also 
\verb:p4.p5: is replaced (we assume that there are fewer than 34 powers of 
\verb:p1.p2:). The power is that this way we can let the program determine 
by itself what constitutes a good substitution scheme for each topology and 
then work out that scheme. It is this kind of AI that is eventually needed 
if we ever want to go even further.

We determine for each variable that needs to be replaced the maximum power.
If the maximum power of x were to be 20, and if there would already be 5 
table elements from another variable we produce the statements

{\footnotesize
\begin{verbatim}
      Fill dotmaptabl(`T',6) = id_(x^17,x^17,x,x1+...+x12);
      Fill dotmaptabl(`T',7) = id_(x^13,x^13,x,x1+...+x12);
      Fill dotmaptabl(`T',8) = id_(x^9,x^9,x,x1+...+x12);
      Fill dotmaptabl(`T',9) = id_(x^5,x^5,x,x1+...+x12);
      Fill dotmaptabl(`T',10) = id_(x^6,x^3*(x1+...+x12)^3,x^5,
                      x^3*(x1+...+x12)^2,x^4,x^3*(x1+...+x12));
      Fill dotmaptabl(`T',11) = id_(x,x1+...+x12);
\end{verbatim}
}

\noindent
in which T defines the topology. We have a procedure that does this 
automatically but it is too unreadable to be shown here even though it 
takes only 17 lines. This way we define a table with statements waiting to 
be executed. If \verb:$M: is the maximum value of all the table elements 
for each topology we can now execute all statements with

{\footnotesize
\begin{verbatim}
  #do i = 1,`$M'
    id  x?{<Topo1>,...,<Topo11>}[n] = x*dotmaptabl(n,`i');
    .sort:dotmap-`i'/`$M';
  #enddo
\end{verbatim}
}

\noindent
assuming that undefined table elements are replaced by one (new option).
This executes the different statements for the different topologies in 
tandem each optimized for its topology and its maximum power and we have the 
optimal number of modules.

%--#] Tables of statements :
%--#[ Expansions :

\section{Expansions}

As many people here have experienced, IBP reductions involving many loops 
can lead to extremely complicated rational polynomials for the 
coefficients. This is the same in Forcer. With the high complexities of 
some of our integrals we have run in polynomials involving $\epsilon^{500}$ 
and worse.

In Mincer we could expand and because Mincer does not have spurious poles, 
six powers in $\epsilon$ were usually enough. We did not manage to make 
Forcer free of spurious poles, and it may even be impossible. In such a 
case the safest solution is to work with (exact) rational polynomials, but 
these can become 'demanding': At N=4 (Mellin moment) for the coefficient 
functions in DIS to four loops already a maximum term size of 140K was not 
enough for some diagrams. The solution is to offer further possibilities: 
changing from exact to expansion to a predefined depth. What is a 
sufficient depth can be extrapolated from lower values of N.

Example: The declaration

{\footnotesize
\begin{verbatim}
    PolyRatfun rat;
\end{verbatim}
}

\noindent
would define exact rational polynomial coefficients in the function rat. 
At a later stage one could switch to

{\footnotesize
\begin{verbatim}
    PolyRatfun rat(expand,ep,14);
\end{verbatim}
}

\noindent
after which \FORM would expand the rational polynomial to 14 terms deep, 
counting from the leading term in $\epsilon$ or $1/\epsilon$. After this it 
would continue in this mode.

There is also a mode in which \FORM can keep just the leading term and give 
it one for its coefficient. This is rather fast and allows one to determine 
what would be a proper depth of expansion.

There is a caveat: For the complicated terms this is much faster, because 
the rational calculus is now replaced by a normal `addition', but since 
most terms have rather simple fractions, the expansion may make them more 
complicated and hence in the case of higher moments this mode could lead to 
slower programs. The optimal way (=when to change mode) to use this is 
still under investigation.

Probably the most efficient would be to have an 'absolute' depth of 
expansion that is based on the topology of the term. As one may imagine, 
this would be extremely complicated to implement. Hence this is just in the 
thinking stage. Suggestions are welcome.

%--#] Expansions :
%--#[ Conclusions :

\section{Conclusions}

We have seen a number of new \FORM\ features to make it easier to manipulate 
diagrams, notations and rewrites. This is not necessarily simple.

The new programs have already been used for some rather involved 
calculations in DIS~\cite{DIS2} that go considerable beyond the 
existing literature.
 
Another driving force behind this is to give \FORM\ more and more 
capabilities that allow the user to write ever smarter programs. Programs 
that can write parts of themselves, conditioned by the problem they are 
solving. The drawback of this is that the creator of such programs will 
have to think a bit more. The payoff is a program that can go beyond what 
was possible before.

Figures were made with Axodraw2~\cite{Axodraw2}.

%--#] Conclusions :
%--#[ Bibliography :

%--#] Bibliography :
\end{document}